\documentclass[12pt]{article}
\usepackage{graphicx}
\usepackage{latexsym}
\usepackage{amssymb}
\begin{document}
\vspace*{0.88truein}
\centerline{\bf EVOLUTION OF SPATIALLY INHOMOGENEOUS ECO-SYSTEMS:}
\vspace*{0.035truein}
\centerline{\bf AN UNIFIED MODEL BASED APPROACH}
\vspace*{0.37truein}
\centerline{\footnotesize AMBARISH KUNWAR}
\baselineskip=12pt
\centerline{\footnotesize\it Department of Physics, Indian Institute of Technology,}
\baselineskip=10pt
\centerline{\footnotesize\it Kanpur 208016, India}
\centerline{\footnotesize\it E-mail: ambarish@iitk.ac.in}

\vspace*{0.225truein}

\vspace*{0.25truein}
\abstract{Recently we have extended our the "unified" model of evolutionary ecology
to incorporate the {\it spatial inhomogeneities} of the eco-system
and the {\it migration} of individual organisms from one patch to another
within the same eco-system. In this paper an extension of our recent
model is investigated so as to describe the {\it migration} and
{\it speciation} in a more realistic way.}{}{}

\vspace*{5pt}

\vspace*{1pt}
\section{INTRODUCTION}		
\vspace*{-0.5pt}
\noindent
The evolution of life has been simulated often. Our {\it unified} model
\cite{csk} describes both "micro"-
evolution over {\it ecological} time scales (i.e. birth, aging and natural
death of individual organisms) and "macro"-evolution over {\it geological}
time scales (i.e., the origination, evolution and extinction of species).
Recently we have extended our unified model
to incorporate the spatial inhomogeneities of the
eco-system and the migration of individual organisms from one patch to
another \cite{skc}. In this paper an extension of our recently developed model
is investigated to describe the migration and speciation in a more realistic
manner.
The paper is organized as follows. Section 2 gives details of
the model. Section 3 describes the main results and conclusions are
drawn in section 4.

\section{THE ECOSYSTEM}
\label{sec2}
\subsection{The Architecture}
\noindent
We model ecosystems on a  square lattice where each lattice site represents
a distinct {\it patch}, the total number of lattice sites being $T_{max}$. A {\it
self-organising} hierarchical foodweb \cite{foodweb,foodweb1} describes the
prey-predator relation at
each lattice site. We assume that species are organised in different trophic
levels of this hierarchal foodweb at each {\it patch}. The {\it hierarchical}
structure of the foodweb captures the well-known fact that fewer
species exist in higher trophic levels for normal ecosystems. The allowed
range of levels, ${\ell}$ at any patch is $1 \leq {\ell} \leq {\ell}_{max}$. We
assume that only one  species occupies the highest level $\ell = 1$. Each
node at level ${\ell}$ leads to $m$ branches at the level ${\ell} + 1$;
therefore the maximum allowed number of nodes in any level ${\ell}$ is $m^{{\ell}-1}$.
Each node represents a niche that can be occupied by at most one species
at a time. If a species occupies the $\nu$-th node in the
${\ell}$-th trophic level of the food web at lattice site $T$ then we denote
its position by the set $ T,{\ell}, \nu$.
\\At any arbitrary instant of time $t$
the total number of species $N(t)$ at any site $T$
can not exceed $N_{max} = (m^{{\ell}_{max}}-1)/(m-1)$, the total number of
nodes. Our model allows $N(t)$ to fluctuate with time over the range
${\ell} \leq N(t) \leq N_{max}$, where ${\ell}$ itself can fluctuate over
the range  $1 \leq {\ell} \leq {\ell}_{max}$. The population of $i$-th species
at any arbitrary instant of time $t$ is given by $n_i(t)$. Competition among the
individuals of the same species for limited availability of resources, other
than food, imposes an upper limit $n_{max}$ on the allowed population of each
species. Thus, the total number of individuals $n(t)$ at any arbitrary site
$T$ at time $t$ is given by  $n(t) = \sum_{i=1}^{N(t)} n_i(t)$ where
$N(t)$ is the number of species at that site. $T_{max}$, ${\ell}_{max}$
$N_{max}$ and $n_{max}$ are time-independent parameters in the model.

\subsection{Prey-predator interactions}
\noindent
The prey-predator interaction between any two species $i$ and $k$ that occupy
two adjacent trophic levels is denoted by $J_{ik}$. In this model $J_{ik}$ can
take three values $+1, -1$ or $0$. The sign of $J_{ik}$ gives the  direction
of nutrient flow, i.e. it is $+1$ if $i$ is predator of $k$, it is
$-1$ if $i$ is prey of $k$. Thus, $J_{ik} = 0$ means that there
is no prey-predator relation between the two species $i$ and $k$.
                                                                                                                             
The elements of matrix $J$ account for the {\it inter}-species as well as
{\it intra}-species interactions. Let $S_i^+$ be the number of all
prey individuals for species $i$ on the lower trophic level, and $S_i^-$
be $m$ times the number of all predator individuals on the higher
trophic level. Since we assume that a predator eats $m$ prey per time interval
(because of larger body size of predators \cite{cohen03}),
$S_i^+$ gives total available food for species $i$, and $S_i^-$ is the
contribution of species $i$ to all predators on the higher level.
                                                                                If $n_i-S_i^+$ is larger than $S_i^-$ then food shortage will be the
dominant cause of premature death of some individuals of species $i$, even if
none of them is killed by any predator. In this way our model  accounts not
only for the inter-species prey-predator interactions but also for the
intra-species interactions arising from the competition of individuals
of species during shortage of food. On the other hand, if $S_i^- > n_i-S_i^+$,
then some organisms will be eliminated from the existing population due to
killing by predators. To capture the {\it starvation deaths and killing by
the predators}, in addition to the natural death due to ageing, a reduction of
the population by
\begin{equation}
C ~~\max(S_i^-,~n_i- S_i^{+})
\label{eq-kill}
\end{equation}
is implemented at every time step, where $n_i$ is the population of
the species $i$ that survives after the natural death. $C$ is a constant
of proportionality. If the reduction of population leads to $n_i \leq 0$,
species $i$ becomes extinct.
\\We assume that the simplest species (bacterium) occupies the lowest trophic
level and always gets enough resources that neither natural death nor
predators can affect their population.
We do not monitor the ageing and death of the the
``bacteria'' occupying the lowest level of the food web; instead,
we assume a constant population of the ``bacteria'' throughout
the evolution.

\subsection{Collective characteristic of species}
\noindent
An arbitrary species $i$, occupying the $\nu$-th node at the ${\ell}$-th
level of any {\it patch} $T$ is {\it collectively} characterized by
\cite{csk}:\\
(i) the {\it minimum reproduction age} $X_{rep}(i)$,\\
(ii) the {\it birth rate} $M(i)$,\\
(iii) the {\it maximum survival age} $X_{max}(i)$. \\
An individual of the species $i$ can give birth to offsprings only
after attaining the age $X_{rep}(i)$. For simplicity, we assume the
reproduction to be {\it asexual}. Whenever an organism of this species
gives birth to offsprings, $M(i)$ of these are born simultaneously. The
{\it maximum survival age} $X_{max}(i)= 100 \times 2^{(1-\ell)/2}$ of a
species depends only on the trophic level occupied by it. None of the
individuals of this species can live longer than $X_{max}(i)$, even if an
individual somehow manages to escape its predators.

\subsection{Birth and natural death}
\noindent
At each time step, each individual organism $\alpha$ of the species
$i$ gives birth {\it asexually} to $M(i)$ offsprings with a probability
$p_b(i,\alpha)$. The {\it time-dependent} birth probability
$p_b(i,\alpha)$ is a product of two factors. One of these two factors,
$(X_{max}-X)/(X_{max}-X_{rep})$, decreases linearly with age, from unity,
attainable at the minimum reproduction age, to zero at the maximum age. The
other factor is a standard Verhulst factor $ 1 - n_i/n_{max}$ which takes
into account the fact that the eco-system can support only a maximum of
$n_{max}$ individual organisms of each species.
                                                                                
Each individual organism, irrespective of its age, can meet its
natural death. However, the probability $p_d$ of this natural death
depends on the age of the individual. In order to mimic age-independent
constant mortality rate in childhood, we assume the probability $p_d$ of
natural death (due to ageing) to be a constant $p_d = \exp[-r(X_{max}-
X_{rep})/M] $, (where $r$ is a small fraction), so long as $X < X_{rep}$.
However, for $X > X_{rep}$, the probability of natural death is assumed to
increase following the Gompertz's law $p_d = \exp[-r(X_{max}- X)/M]$.
\cite{austad}

\subsection{Mutations}
\noindent
\subsubsection{Effect of mutation on species}
$X_{rep}$ and $M$ of  each of the species of the eco-system randomly increases
or decreases, with equal probability, by unity with probability $p_{mut}$ per
unit time due to mutation. $X_{rep}$ is restricted to remain in the interval
from $1$ to $X_{max}$, and $M > 0$.

\subsubsection{Effect of mutation on foodweb}
\noindent
In order to reduce the computational requirements it has been assumed that food
habits of all species occupying similar niche on different patch (i.e., all nodes with identical set of index $\ell$ and $\nu$ irrespective of $T$) changes
simultaneously due to mutation.
To capture this; with the same probability $p_{mut}$
per unit time, one of the links $J$ from prey and one of the links $J$ to
predators is re-adjusted \cite{sole}.
If the link $J_{ij}$ to the species $i$ from a {\it higher} level species $j$
is non-zero, it is assigned a new value of $J_{ij} = J_{ji} = 0$. On the
other hand, if the link $J_{ik}$ to a species $i$ from a {\it lower}
species $k$ is zero, the new values assigned are $J_{ik} = 1, J_{ki} = -1$.
These readjustment of the prey-predator interactions is motivated by the fact
that each species tries to minimize its predators but, at the same time, looks
for new prey (food).
                                                                                
\subsection{Speciation}
\noindent
The niches (nodes) left empty because of
extinction are re-filled by new species, with probability $p_{sp}$
per unit time \cite{orr}. All the simultaneously re-filled nodes in a
trophic level of foodweb originate from {\it one common ancestor}
which is picked up randomly from among the non-extinct species
at the same trophic level. The characteristic parameters $X_{rep}$, $M$ of
each of the new species are identical to their ancestor unlike earlier
models \cite{csk,chow2,chow3,chow4,skc} but can mutate later. However, occasionally, all the niches at a level may lie
vacant.
Under such circumstances, all these vacant nodes are to be filled
by a non-extinct species occupying the closest {\it lower} populated
level.

\subsection{Emergence of a new trophic level}
\noindent
If the total biomass of the surviving individuals in the whole ecosystem
is below the preassigned maximum possible value, a new trophic level is created
, with probability $p_{lev}$, i.e. $\ell$ is increased by one. To
prevent the number of levels to increase towards $\ell_{max}$, a new level could be
created only if the total population on that lattice site was at most
$2 n_{max} = 200$. The actual number of occupied levels can fluctuate with
time depending on extinction and speciation. The total number of levels,
which determines the lengths of the food chains, depends on several
factors, including the available bio-mass \cite{post}.
                                                                                
\subsection{Migration}
\noindent
The individuals of a species can migrate to a neighboring lattice site with
probability $p_{mig}$ per unit time, if the
population at the corresponding niche (i.e. at the same level and at the same
node) of the neighboring site is zero for the same species on the same level.
The population at the old site diminishes by a random fraction and exactly
the same number shows up at the neighboring site. The age of these
immigrants do not change on migration but at their new habitat they
continue to age as usual (unlike our recent model \cite{skc}).

\subsection{Initial conditions and update rules}
\noindent
The requirements of computational resources increase rapidly with
increasing $T_{max}$, $\ell_{max}$, $m$, $n_{max} $. Therefore, in almost all
simulations values of these parameters were $T_{max}=169$, $\ell_{max} = 10$,
$m=2$, $n_{max}=100$.
Simulations always began with a random initial condition
where $M = 1$ for all species, with only three levels in the foodweb.
Since larger species occupy the higher tropic levels and are expected to
live longer than those at lower levels, we assigned $X_{max} = 100, 71,
50, 35, 25, 18, 12, 9, 6, 4$ to the species at level ${\ell} = 1, 2, 3, 4,
5, 6, 7, 8, 9, 10$, respectively. Initially, $X_{rep}$ was assigned
randomly between $1$ and $X_{max}$, the population randomly between $1$
and $n_{max}/2$. The ages of the individuals were chosen randomly between
$1$ and the $X_{max}$ of the corresponding species. We allowed the
eco-system to evolve for  $t_{warm}$ time steps before we started
collecting data from it and the data were collected for subsequent $t = 5
t_{warm}$ time steps.
\\The state of the system is updated in discrete time steps where each
step consists of a sequence of seven stages:\\

\noindent {\it I- Birth}
                                                                                
\noindent {\it II- Natural death}
                                                                                
\noindent {\it III- Mutation}
                                                                                
\noindent {\it IV- Starvation death and killing by prey}
                                                                                
\noindent {\it V- Speciation}
                                                                                
\noindent {\it VI- Emergence of new trophic level}
                                                                                
\noindent {\it VII- Migration}

\section{RESULTS}
\noindent
\label{result}

\begin{figure}[htbp]
\begin{center}
\includegraphics[angle=-90,width=0.50\columnwidth]{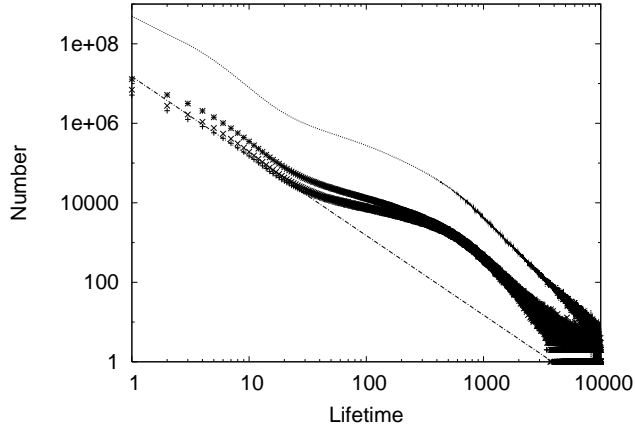}
\end{center}
\caption{Histogram for the lifetimes
of species, with migration on a
$13 \times 13$ square lattice, with
observation time $t = 10^4$($+$), $10^5$($\times$), $10^6$($\ast$) summing over
$10^2, 10, 1$ sample
respectively. One run with $13 \times 13$ lattice for observation time
$t = 10^7$ is shown with dots. The line with slope -2 corresponds to a
power-law distribution predicted by several models. The common parameters for
all the plots are $m=2$, $n_{max} = 100$, $p_{sp} = 0.1, p_{mut} = 0.01,
p_{lev} = 0.001, p_{mig}=0.5$, $C = 0.05$, $r = 0.05$.}
\label{fig-1}
\end{figure}

\subsection{Lifetime distribution}
\noindent
The distributions of lifetimes of species are plotted in Fig.\ref{fig-1}
for different sets of values of the parameters. The Power-law with an
exponent of -2, claimed by several simple models based on macro-evolution
\cite{drossel,newman},
holds approximately only over the shorter lifetime regime; for longer lifetimes
a strong deviation from power-law is observed. Since available fossil data
have various known difficulties, it is questionable whether real extinctions
follow power laws and, if so, over how many orders of magnitude.

\begin{figure}[htbp]
\begin{center}
\includegraphics[angle=-90,width=0.50\columnwidth]{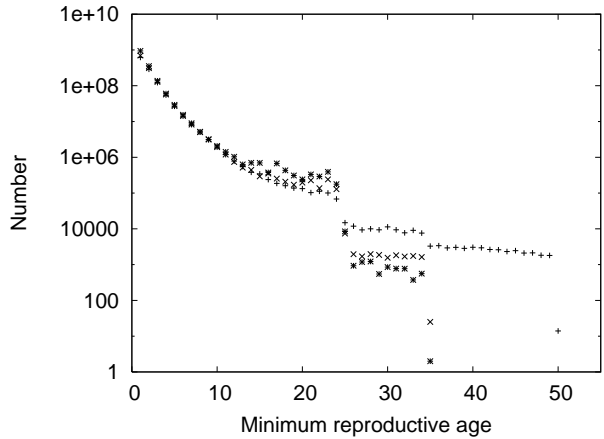}
\end{center}
\caption{Histogram for the minimum reproductive age
of species, with migration on a
$13 \times 13$ square lattice, with
observation time  $t = 10^4$($+$), $10^5$($\times$), $10^6$($\ast$) summing
over $10^2, 10, 1$ sample
respectively. The values of the common parameters for all the plots are
identical to those used in Fig. 1}
\label{fig-2}
\end{figure}

\subsection{Distribution of minimum reproductive ages}
\noindent
The distributions of minimum reproductive age $X_{rep}$ of the species for
different set of values of parameters have been show in Fig.\ref{fig-2}.
The distribution of minimum reproductive age broadens in the early stages of
macro-evolution but it shrinks in the late stages where it appears to take
a stationary form where largest value of minimum reproduction age is around 35.

\begin{figure}[htbp]
\begin{center}
\includegraphics[angle=-90,width=0.50\columnwidth]{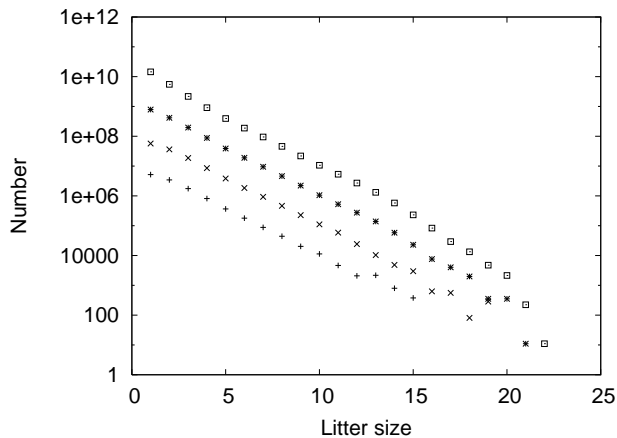}
\end{center}
\caption{Histogram for the litter size, with migration on a
$13 \times 13$ square lattice, for observation time
$t=10^4$($+$), $10^5$($\times$), $10^6$($\ast$) and $10^7$($\square$) for $1$
sample only. The values of the common parameters for all the plots are
identical to those used in Fig. \ref{fig-1}}
\label{fig-3}
\end{figure}

\subsection{Distribution of litter size}
\noindent
Fig.\ref{fig-3} shows the litter size distribution for different set of model
parameter values. The litter size distribution is roughly exponential.

\begin{figure}[htbp]
\begin{center}
\includegraphics[angle=-90,width=0.50\columnwidth]{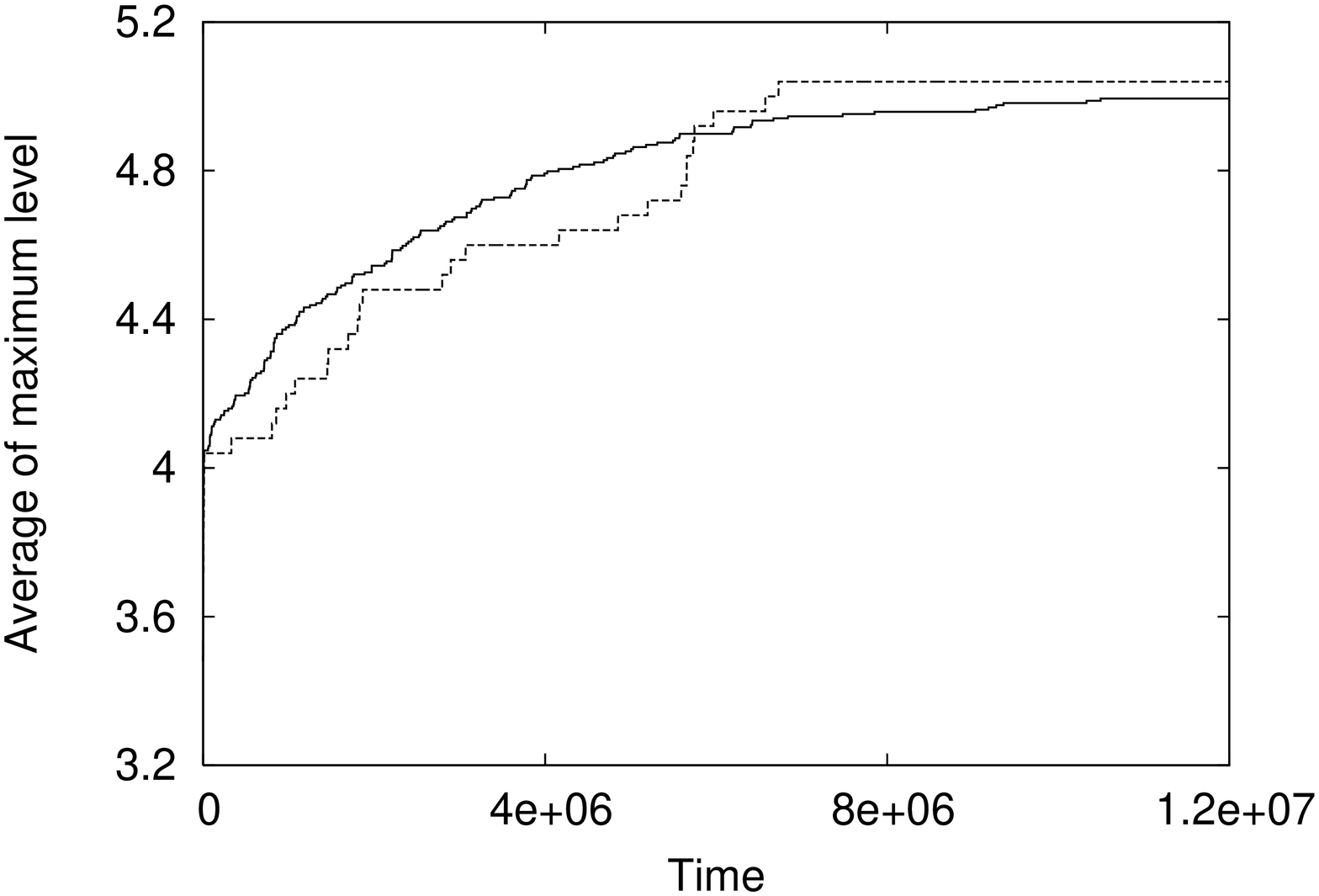}
\includegraphics[angle=-90,width=0.50\columnwidth]{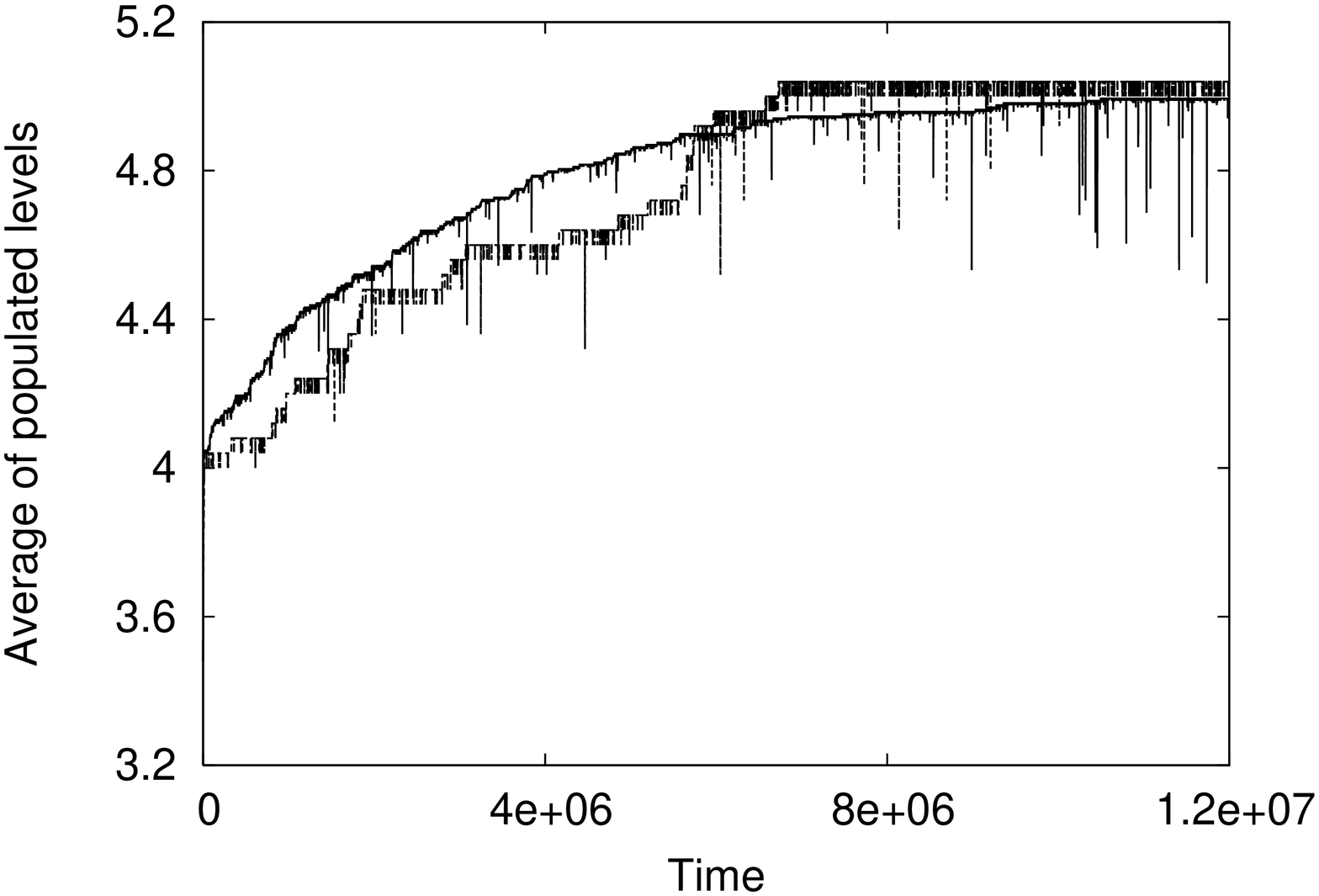}
\end{center}
\caption{Average of maximum level as function of time (upper plot). Average of
number of populated levels as a function of time (lower plot). $T_{max} =
13 \times 13$ (with solid line) and $T_{max} = 5 \times 5$ (with dashed line),
otherwise the common parameters for all the
plots are as in Fig. \ref{fig-1}}
\label{fig-4}
\end{figure}
                                                                                
\subsection{The number of populated levels}
\noindent
At some lattice sites all the species in a trophic level may become extinct
occasionally due to randomness in the evolutionary process and all niches
in that level may lie vacant (this may happen
for all levels except the lowest one occupied by bacteria). Variation of
average of maximum level (quantity
$ \sum_{i=1}^{T_{max}} l_i(t)/T_{max}$ where $l_i(t)$ is the maximum number of level
at site $i$ at time $t$) and  average of number of populated levels
(quantity $ \sum_{i=1}^{T_{max}} p_i(t)/T_{max}$ where $p_i(t)$ is the number of those trophic levels at site $i$ at time $t$ in which at least one niche is
occupied by a non-extinct species) with time is shown for a single run in
Fig.\ref{fig-4}.
Fig.\ref{fig-4} upper plot clearly shows a growing bio-diversity during
early stages of the evolution. Lower plot shows that
the actual number of occupied levels and hence the total number of species in
the eco-system keep fluctuating at all stages of
evolution due to extinction and speciation.

\begin{figure}[htbp]
\begin{center}
\includegraphics[angle=-90,width=0.50\columnwidth]{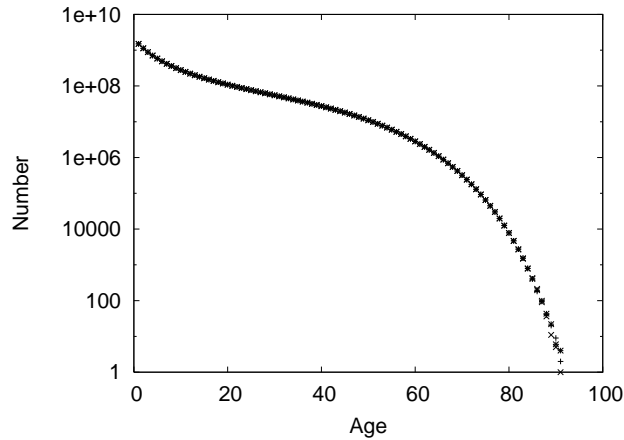}
\end{center}
\caption{Age distribution of the individuals occupying the highest
level, with migration on a
$13 \times 13$ square lattice, for $100$ samples and $10^4$ time steps
($+$), $10$ samples and $10^5$ time steps($\times$), $1$ sample and $10^6$
time steps($\ast$). The values of the common parameters for all the plots are
identical to those used in Fig. \ref{fig-1}}
\label{fig-5}
\end{figure}

\begin{figure}[htbp]
\begin{center}
\includegraphics[angle=-90,width=0.50\columnwidth]{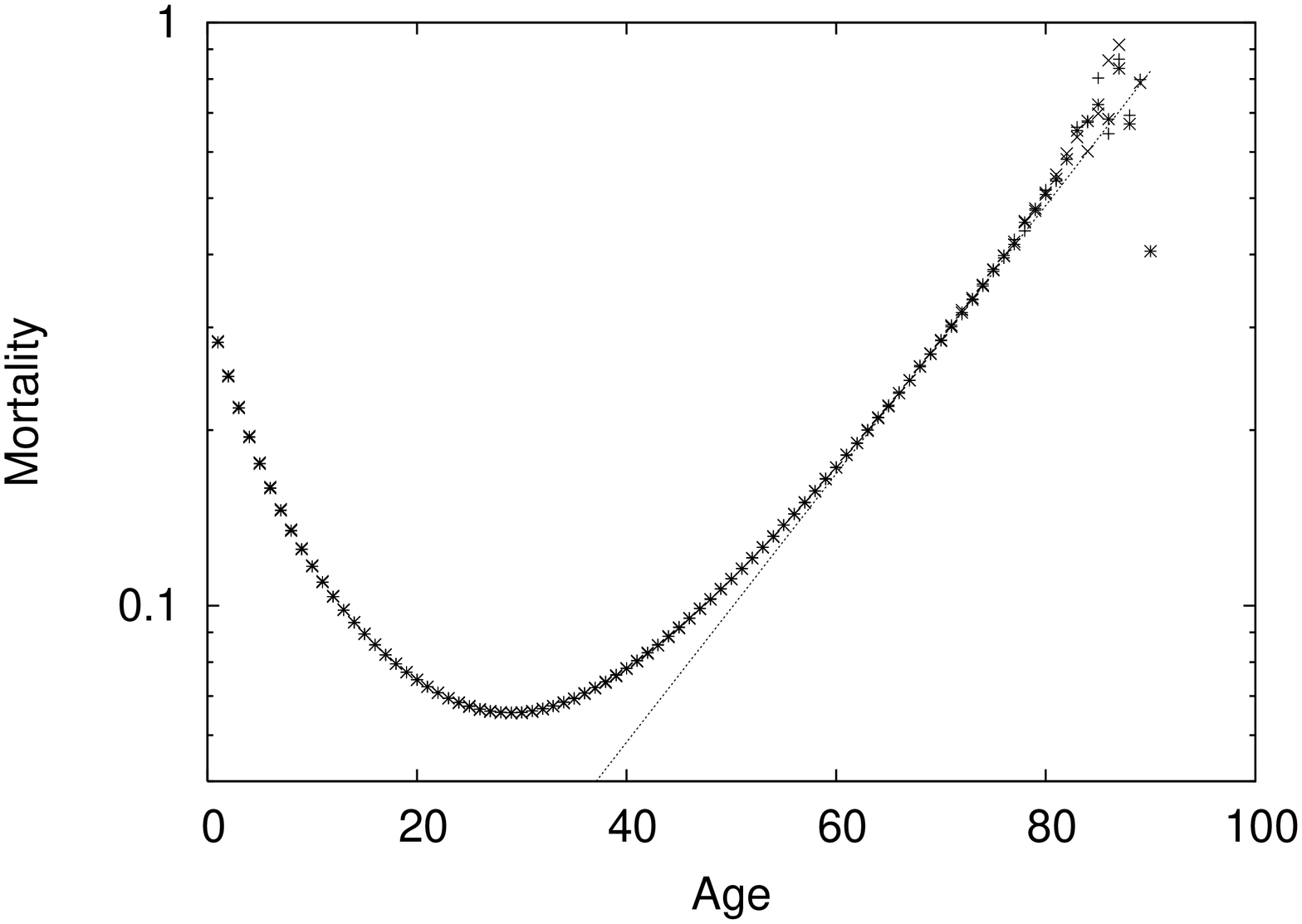}
\end{center}
\caption{
Mortality rate extracted from the age distribution shown in Fig. \ref{fig-5}.
Gompertz's law corresponds to straight line.}
\label{fig-6}
\end{figure}

\subsection{Age distribution and Mortality rate}
\noindent
Fig.\ref{fig-5} shows the age distribution of the individuals occupying
the highest trophic level $\ell=1$. The mortality rate defined by the
relation -d ($ln$[Age distribution])/d(age) has been extracted from the age                                                                                
distribution of the individuals at the highest trophic level is shown in
Fig.\ref{fig-6}. The mortality curve is consistent with the usual census data
where mortality rate goes through a minimum in the childhood and increases
exponentially with age in adults.

\begin{figure}[htbp]
\begin{center}
\includegraphics[angle=-90,width=0.50\columnwidth]{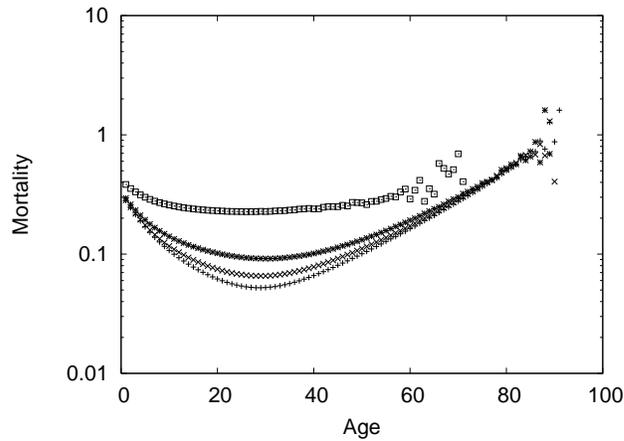}
\end{center}
\caption{variation of mortality for different values of parameter $C$.
$C=0.02$($+$), $C=0.05$($\times$), $C=0.1$($\ast$), $C=0.2$ ($\square$).
The values of the common parameters for all the plots are
identical to those used in Fig. \ref{fig-1}}
\label{fig-7}
\end{figure}

\subsection{Natural death versus death due to predators(starvation)}
\noindent
Fig.\ref{fig-7} shows variation of mortality rate for different values
of $C$. With increasing values of $C$, death due to predators/hunger
starts dominating over the genetic(natural) death and mortality function
becomes almost independent of age. However, when genetic death dominates
the mortality function obeys Gompertz's law.
Branching ratio of 2 gave highest variation of mortality in simulations but
the range over which mortality varied was identical for higher branching
ratio 3 and 4 (not shown).

\begin{figure}[htbp]
\begin{center}
\includegraphics[angle=-90,width=0.50\columnwidth]{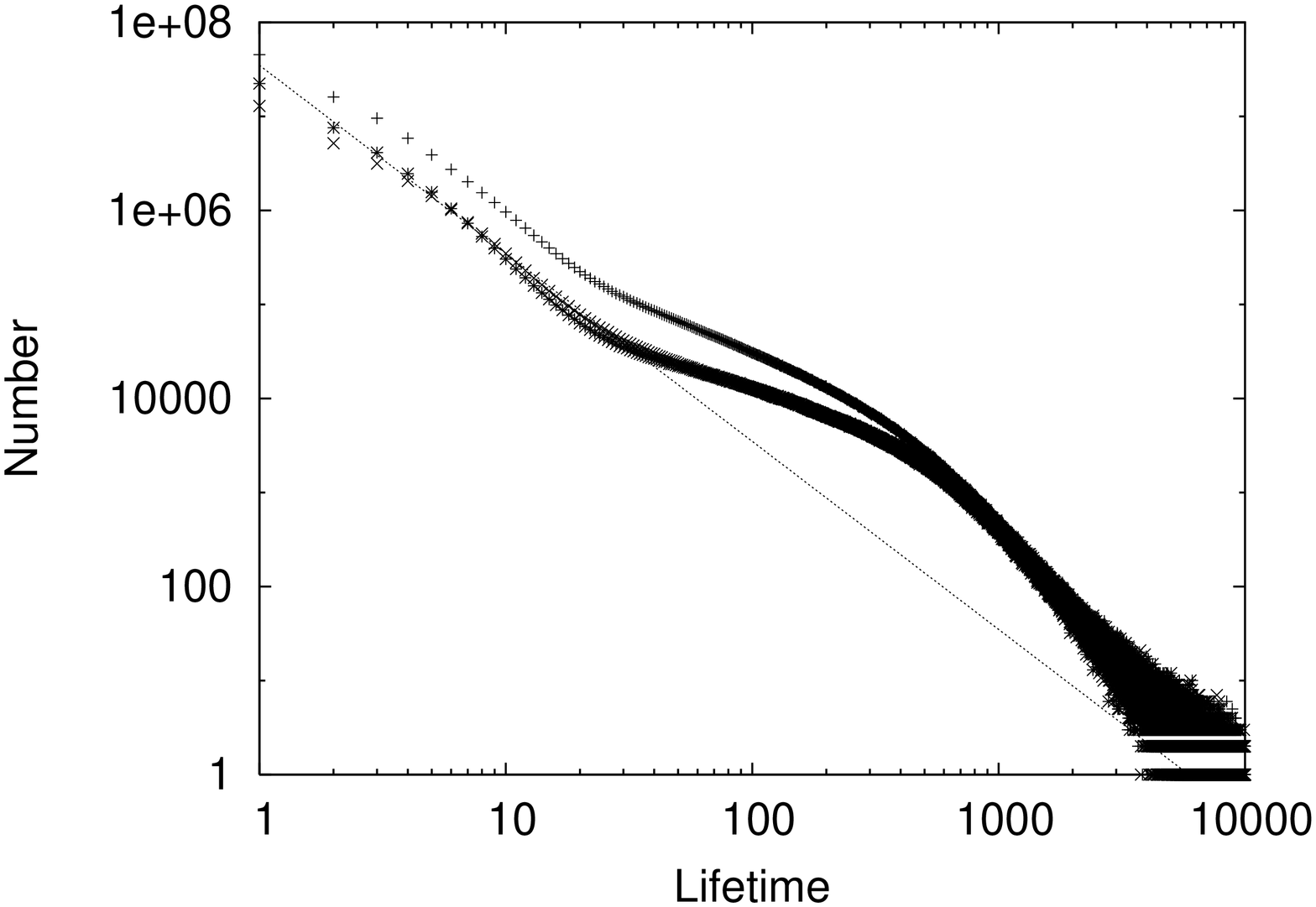}
\includegraphics[angle=-90,width=0.50\columnwidth]{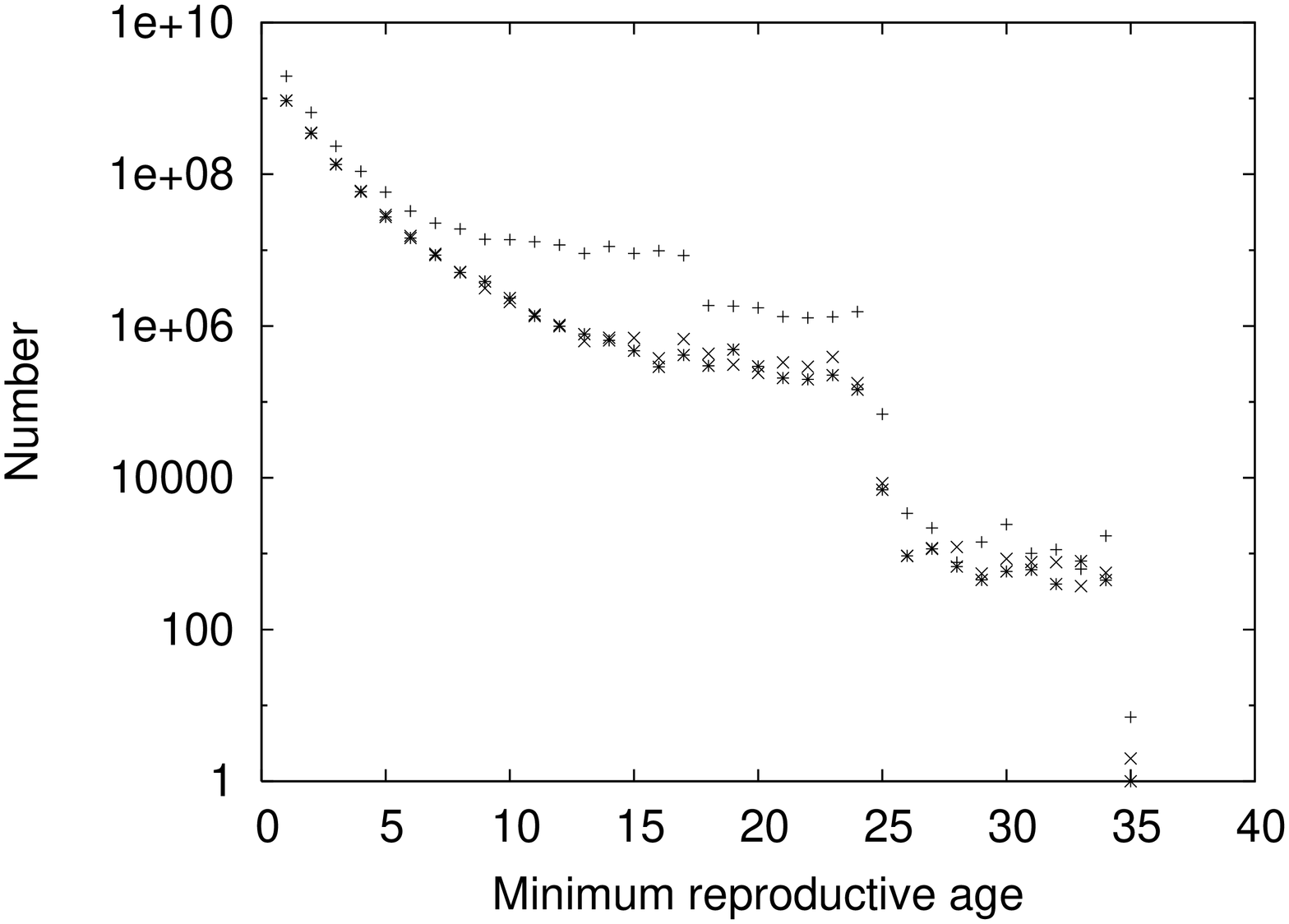}
\end{center}
\caption{Lifetime distribution and Minimum reproductive age distribution
for migration probabilities $p_{mig}=0.0$($+$), $0.5$($\times$), $1.0$($\ast$)
with $1$ sample and $10^6$ time steps. The common parameters for all the
plots are $T_{max} = 13 \times 13, m=2$, $n_{max} = 100$, $p_{sp} = 0.1,
p_{mut} = 0.01, p_{lev} = 0.001 $, $C = 0.05$, $r = 0.05$.}
\label{fig-8}
\end{figure}

\section{SUMMARY AND CONCLUSION}
\noindent
\label{summary}
In this paper, the "unified" model of eco-system, which incorporates the
spatial inhomogeneities of the eco-system and
the migration of individual from one patch to another within the same
eco-system has been extended to describe the migration and speciation in
a more realistic manner. In this model the age of immigrants do not change
on migration and they continue to age as usual at their new habitat.
Moreover, in this model all empty niches initially filled (occupied) by one common
ancestor during speciation lead to different species (due to random mutations occurring
independently for each filled niche) in due course of evolution. Qualitative
features of lifetime distributions and minimum reproductive age distributions
are barely affected by migration (see Fig.\ref{fig-8}).
\\We presented a systematic study of the dependence on input parameters.
The model presented here is capable of {\it self-organization}; the total number of species, prey-predator interactions in the foodweb, number of trophic
levels in the foodweb and the collective characteristics of the species
are emergent properties of the self-organizing dynamics of the model.

\section{Acknowledgements}
\noindent
AK thanks D. Stauffer and D. Chowdhury for
useful discussions and DFG/Sta130 for partial support.

\end{document}